\documentclass[11pt]{article}
\usepackage{a4}
\usepackage{amsmath}
\usepackage{amsfonts}
\usepackage{latexsym}
\usepackage{epsfig}
\usepackage{color}
\if@twoside\oddsidemargin25mm
\else
\newcommand{\be}{\begin{equation}}
\newcommand{\ee}{\end{equation}}
\newcommand{\bea}{\begin{eqnarray}}
\newcommand{\eea}{\end{eqnarray}}

\renewcommand{\d}{\mathrm{d}}

\newcommand{\Ra}{\Rightarrow}

\DeclareMathSymbol{\mg}{\mathrel}{symbols}{"1D}

\renewcommand{\gg}{\gamma}
\newcommand{\gd}{\delta}

\newcommand{\gf}{\phi}

\newcommand{\gx}{\xi}
\newcommand{\gm}{\mu}

\newcommand{\gl}{\lambda}

\newcommand{\gp}{\pi}
\newcommand{\gps}{\psi}
\newcommand{\get}{\eta}

\newcommand{\gG}{\Gamma}

\newcommand{\gF}{\Phi}

\newcommand{\gL}{\Lambda}

\newcommand{\gO}{\Omega}

\newcommand{\tD}{{\tilde D}}

\newcommand{\tr}{\mbox{tr}}

\newcommand{\ra}{\rightarrow}

\newcommand{\der}{\partial}

\newcommand{\nit}{\noindent}

\newcommand{\dsp}{\displaystyle}

\newcommand{\labl}[1]{\label{#1}}
\newcommand{\half}{\frac 12 }

\newcommand{\beq}{\begin{equation}}
\newcommand{\eeq}{\end{equation}}
\newcommand{\barr}{\begin{array}}
\newcommand{\earr}{\end{array}}
\newcommand{\equ}[1]{\begin{equation} #1 \end{equation}}

\newcommand{\tabu}[2]{\begin{tabular}{#1} #2 \end{tabular}}
\newcommand{\arry}[2]{\begin{array}{#1} #2 \end{array}}

\newcounter{oldcounter}

\newcommand{\bgf}{{\bar\phi}}

\newcommand{\Natr}{\mathbb{N}}
\newcommand{\Intr}{\mathbb{Z}}

\begin{document}

\begin{flushright}
\uppercase{\mbox{UVIC-TH/02-03}}, \\ 
hep-th/0210338
\end{flushright}
\vskip 2 cm
\begin{center}
{\Large {\bf 
Localization and anomalies on orbifolds
} 
}
\\[0pt]

\bigskip
\bigskip {\large
{\bf S.\ Groot Nibbelink$^{a,}$\footnote{
{{ {\ {\ {\ E-mail: grootnib@uvic.ca}}}}}}}, 
{\bf H.P.\ Nilles$^{b,}$\footnote{
{{ {\ {\ {\ E-mail: nilles@th.physik.uni-bonn.de}}}}}}},
\bigskip }\\[0pt]
\vspace{0.23cm}
${}^a${\it  
Department of Physics and Astronomy, University of Victoria,
} \\
{\it 
PO Box 3055 STN CSC, Victoria, BC, V8W 3P6 Canada 
}\\
{\it 
CITA National Fellow
}\\
\vspace{0.23cm}
${}^b$ {\it  Physikalisches Institut der Universit\"at Bonn,} \\
{\it Nussallee 12, 53115 Bonn, Germany.}\\
\bigskip
\vspace{3.4cm} {\bf Abstract}

\end{center}

In this talk we discus some properties of supersymmetric 
theories on orbifolds in five dimensions. The structure of
FI--tadpoles may lead to (strong) localization of charged bulk
scalars. Orbifold theories may suffer from various kinds of
anomalies. The parity anomaly may render the construction of the
orbifold theory ill--defined. The gauge anomaly on the orbifold 
are localized at the fixed points, which can sometimes be canceled by
a Chern--Simons term.

\vfill
{\em 
\nit
Based on talks given by the authors at: 
\\[1ex]
SUSY'02, ``the 10th International Conference on Supersymmetry and
Unification of Fundamental Interactions'',  DESY, Hamburg, Germany,
17-23 June 2002,  and 
\\[.5ex]
``The 1st International
Conference on String Phenomenology'', Oxford, July 6 - 11, 2002.
}

\newpage 

\section{Introduction and summary}

Recently there has been a large interest in field theories with 
extra dimensions. When these extra dimensions have boundaries, 
one has to consider field theories with fields living both in the 
bulk and on these boundaries. As those theories are presumably some 
sort of low--energy description of string or M--theory, they often
contain some remnant of supersymmetry. At the same time one would hope
that they allow for phenomenology which resembles physics of the 
(supersymmetric) standard model. 

A central question in the discussion of these field-theoretic
orbifolds concerns the stability with respect to ultraviolet
effects as, for example, quadratic divergences of scalar mass terms.
One way to insure stability in this respect would be the
consideration of supersymmetry \cite{Witten:nf}. In the presence of $U(1)$
gauge groups, however, supersymmtery is not enough as there might
appear quadratically divergent Fayet-Iliopoulos (FI) terms even 
within the supersymmetric context. To obtain a stable theory such 
FI-terms have to be cancelled by a specific choice of the $U(1)$
charges of scalar fields. Higher order corrections are absent due
to a nonrenormalization theorem\cite{Fischler:1981zk}. The stability question
of field theoretic orbifolds has been discussed extensively in the
literature; see ref.\ \cite{Ghilencea:2002jd} for a review and references.

This discussion is quite relevant for phenomenological considerations
as the standard model contains the $U(1)$ gauge group of hypercharge.
Quadratically divergent FI-tadpoles will appear if the sum of the
hypercharges of the scalar fields does not vanish. A theory with a
single Higgs multiplet (as in the standard model) will thus
generically suffer from an ultraviolet instability. The simplest
way to avoid this problem is the introduction of a second Higgs
multiplet with opposite hypercharge (as e.g. in the minimal
supersymmetric standard model).

In higher dimensional theories with bulk and brane fields the above
discussion will become even more complex as localized FI-tadpoles
(at some boundary or fixed point) might appear. In the present talk we
shall elaborate on these complications and discuss the physical 
consequences of localized anomalies and FI-tadpoles.
For simplicity we consider supersymmetric 
gauge theory in five dimensions compactified on $S^1/\Intr_2$, coupled
to hyper multiplets in the bulk and chiral multiplets on the
boundaries. Before discussing these aspects in more detail, let us
summarize the results that are reviewed in this talk:
 
The shape of the FI--terms over the fifth dimension leads 
to intriguing physical effects: it can cause the localization of the
zero modes of the bulk hyper multiplets to the branes. 
Another important issue in field theory is the structure of
anomalies. For orbifolds the anomaly structure can be quite rich: 
the gauge anomalies tend to be localized at the branes, and their 
cancellation may involve some contributions of a five dimensional 
Chern--Simons term. Another type of anomaly, the so--called parity 
anomaly, may lead to difficulties in defining the orbifold theory 
in the first place.

\section{The setup of the 5D supersymmetric orbifold}

We consider a five dimensional bulk on $S^1/\Intr_2$ with 
one vector multiplet $V$ and a set of hyper
multiplets $H = \{H^b, b = 1, \ldots, n\}$. 
The vector multiplet $V = (A_M, \gl, \gF, \vec D)$ contains a 
five dimensional vector field $A_M$, a Dirac gaugino $\gl$, 
a real scalar $\gF$, and a triplet of auxiliary fields 
$\vec D$ in the off--shell formulation. 
The hyper multiplets $H = (\gf_+, \gf_-, \gps_{+L}, \gps_{-L}, f_+,
f_-)$ consist of two complex scalars $\gf_+, \gf_-$,  
called the hyperons, two sets of chiral spinors  $\gps_{+L},
\gps_{-L}$, the so--called hyperinos, and complex auxiliary scalars 
$f_+, f_-$.  They are charged under the $U(1)$ gauge field, with
charge operator $q$. The orbifolding leads to the following parity
assignments 
\[
V:~
\arry{l|c|c|c|c|c|c|c|}{
\mbox{state} &A_\gm & A_5 & \gF & \gl_{\pm L} & \gl_{\pm R} & D_3 & D_{1,2}
\\\hline
\mbox{parity} & + & - & - & \pm & \pm & + & -}
\qquad
H:~ \arry{l|c|c|c|c|}{
\mbox{state} &\gf_\pm & \gps_{\pm L} & \gps_{\pm R} & f_\pm
\\\hline
\mbox{parity} &\pm & \pm & \pm & \pm }
\]

From the five dimensional supersymmetry transformations, one can obtain 
the unbroken four dimensional supersymmetry transformation with
Majorana parameter $\get_+$, which exists at the
orbifold fixed points. This implies \cite{Mirabelli} that the four
dimensional vector multiplet on the branes is given by 
$V| = (A_\gm, \gl_+, \tD_3)$ with ``modified'' auxiliary field 
$\tD_3 = D_3 - \der_y \gF$. 
In addition we allow for an arbitrary number of chiral multiplets
$C_0 = (\gf_0, \gps_{0L}, \tilde f_0)$ and
$C_\gp = (\gf_\gp, \gps_{\gp L}, \tilde f_\gp)$ 
on the branes $y=0$ and $y=\gp R$.

\section{Fayet--Iliopoulos tadpoles and localization of bulk fields}

Now we turn to the first central topic of this talk: the 
structure of the one loop induced FI--tadpoles. 
As $\tD_3$ not $D_3$ is the relevant auxiliary field for the vector
multiplet at the branes, the one-loop FI-terms due to brane chiral 
multiplets, $C_0$ or $C_\gp$, are proportional to $\tD_3$
\equ{
\raisebox{-10mm}{\scalebox{.9}{\mbox{\input{FIbrane.pstex_t}}}} 
~
\gx_{branes}(y) = 
 g_5 {\dsp \frac {\gL^2}{16 \gp^2} }
\dsp 
\sum_{I = 0, \gp}
\gd(y - I R)  \tr(q_I).
}
The situation for tadpoles due to the bulk fields is more subtle. 
In ref. \cite{Ghilencea:2001bw} it was shown that hyper multiplets
may lead to a quadratically divergent zero mode FI--term, and in ref.\
\cite{Scrucca:2002eb} it was argued that the counter term for $D_3$ is
located at the branes. This localization on the branes implies that
there exists tadpole for the $\der_y \gF$ as well. 
In ref. \cite{GrootNibbelink:2002wv,GrootNibbelink:2002xx} 
it was shown that the bulk hyperinos $\psi$ give rise to a tadpole for
this operator. The bulk induced tadpoles involve a double derivative 
acting on the brane delta functions: 
\equ{
\raisebox{-10mm}{\scalebox{.9}{\mbox{\input{FIbulk.pstex_t}}}}
\,
\gx_{bulk}(y) = g_5 \frac{\tr(q)}{2}\!
\left(\!\!
\frac{\gL^2}{16 \gp^2} \!+\!
\frac{\ln \gL^2}{16 \gp^2} \frac {\der_y^2}4 
\!
\right)\!\!
\Bigl[ \gd(y) + \gd(y - \gp R) \Bigr].
}
The derivative $\der_y \gF$ arises because of the Kaluza--Klein mass
coupling of the hyperinos to $\gF$, which is dictated by five dimensional
supersymmetry. Combining both the brane and bulk contributions to 
the FI--terms gives 
\equ{
\gx(y) = \sum_{I=0,\gp}(\gx_I + \gx_I'' \der_y^2 )
\gd(y - IR),
}\equ{
\gx_I = g_5 \frac{\gL^2}{16 \gp^2}
\Bigl(
\half \tr(q) + \tr (q_I)
\Bigr),
\quad
\gx_I^{\prime\prime} = \frac 14 g_5 \frac{\ln \gL^2}{16 \gp^2}
\half \tr(q).
}

\begin{figure}
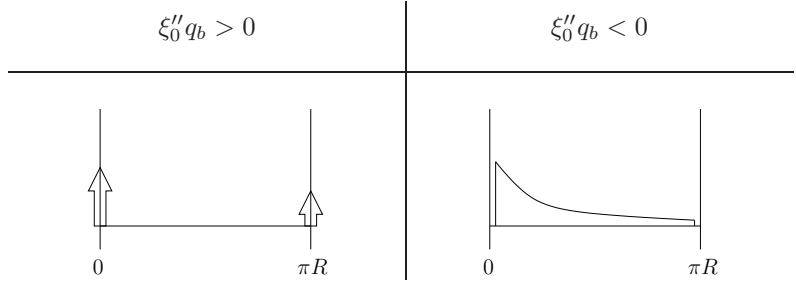

{\small 
\begin{center} 
\renewcommand{\arraystretch}{1.5}
\tabu{ c | c }{
 $\gx_0'' q_b > 0$ & $\gx_0'' q_b < 0$ 
\\[2ex] \hline &\\[-1ex]
~~~~~~
\raisebox{-2mm}{\scalebox{.7}{\mbox{\input{loc0pi2.pstex_t}}}}
~~~~~~
& 
~~~~~~
\raisebox{-2mm}{\scalebox{.7}{\mbox{\input{exp0pi.pstex_t}}}}
~~~~~~
%\\[-1ex] &\\ \hline 
}
\end{center}}
\caption{The two basic shapes (eqs.\ \eqref{wave>} and \eqref{wave<})
of the zero mode with charge $q_b$ are displayed for a finite value of
the cut-off $\gL$. Delta function localizations, denoted by the arrows. 
}
\label{localPosibilitiesCutFinite}
\end{figure}

The second main issue of this talk is
the question how these FI--terms can cause
an instability that finally leads to localization of bulk 
hyper multiplet fields.  
For this we first investigate the background for $\gF$ in the presence
of  the FI--tadpoles. Its (BPS) field equation, that respects four
dimensional supersymmetry, is given by  
\equ{
D_3 = \der_y
\gF = g_5( \gf_+^\dag q \gf_+ - \gf_-^\dag q \gf_-)
  + \sum_{I= 0, \gp} \gd(y -\gp R) g_5 \gf_I^\dag q_I \gf_I 
+ \gx(y).
\labl{BPS1}
}
There are supersymmetric vacua which do not spontaneously break the
gauge symmetry (all charged scalars vanish in the vacuum) 
if the following integrability condition is satisfied 
\equ{
 0 =  \int_0^{\gp R}  \!\!\!\! \d y\, \der_y \langle \gF \rangle 
= \half \gx_0 + \half \gx_\gp 
~~ \Rightarrow ~~
\tr (q) + \tr (q_0) + \tr (q_\gp) = 0.
\labl{vanishingFI}
}
Note that this condition is identical to the requirement that the mixed
$U(1)$ gauge gravitational anomaly is absent in the effective four dimensional
theory. Therefore, from now we always assume that 
$\gx_\gp = - \gx_0$. 
Even with this requirement fulfilled, the shape of the background 
expectation value of $\langle \gF \rangle$ is non--trivial. Its
integral, for example, takes the form 
\equ{
\int_0^{y} \d y\, \left< \gF\right> (y) = \half \gx_0 
\Bigl( \gp R - | y - \gp R| \Bigr) + \gx_0^{\prime\prime} 
\Bigl( \gd(y) + \gd(y - \gp R) \Bigr). 
\labl{intgF}
}
This affects the shape of the zero mode in a dramatic way. 
The shape of the zero mode $\gf^b_{0+}$ of an even bulk field with
charge $q_b$ is given by  
\equ{
\der_y \gf^b_{0+} -  g_5 q_b \left< \gF\right> \gf^b_{0+} = 0 
~~\Ra ~~
\gf^b_{0+}(y) = \exp\left\{  g_5 q_b \int_0^y \d y \left< \gF \right>  \right\} 
\bgf^b_{0+},
\labl{zeroMode}
}
where the integral has been evaluated in \eqref{intgF}. The
remaining $N=1$ four dimensional supersymmetry implies that the 
zero mode wave functions of the even bulk scalar $\gf^b_{0+}$ and the
chiral fermion $\gps^b_{0+L}$ with charges $q_b$ are identical. 
The delta functions in that expression requires some form of
regularization that takes the normalization 
\(
\int_0^{\gp R} \d y\,  |\gf_{0+}(y)|^2 = 1
\)
into account. Skipping the computational details (which can be found in 
ref.\cite{GrootNibbelink:2002xx}) we find that the shape of the zero mode 
crucially depends on the sign of the product of $\gx_0'' q_b$
(the special case $\gx_0'' = 0$ has be studied in 
refs.\cite{Kaplan:2001ga,Arkani-Hamed:2001is}): 
\equ{
(\gf^b_{0+})^2(y) = 
\frac {2 e^{g_5 q_b \gx_0 \, y}}{1 + e^{g_5 q_b \gx_0 \, \gp R}}
\left[
\gd(y) + \gd(y-\gp R)
\right], 
\qquad \quad 
\gx_0^{\prime\prime} q_b > 0; 
\labl{wave>}
}
\equ{
(\gf^b_{0+})^2(y) = 
\frac {g_5 q_b \gx_0\,  e^{g_5 q_b \gx_0 \, y}}{e^{g_5 q_b \gx_0 \, \gp R}-1}
\left\{\arry{ll}{
1 & 0 < y < \gp R, \\[2ex]
0 & y = 0, \gp R,
}\right. 
\qquad \qquad 
\gx_0^{\prime\prime} q_b < 0.
\labl{wave<}
}
Hence, for $\gx_0^{\prime\prime} q_b > 0$  
the zero mode has the delta function support on the two 
fixed points, but the height at these two fixed points is not the
same: while, for $\gx_0^{\prime\prime} q_b < 0$ the zero
mode vanishes at both branes identically, but has an exponential
behavior on the open interval $]0, \gp R[$. 
In both cases, displayed in figures \ref{localPosibilitiesCutFinite}, 
the value of $|\gx_0^{\prime\prime}|$ does not appear anymore; 
it has been absorbed in the regularization of the 
delta functions when implementing the normalization of the modes.
The shapes of the zero modes in the limit of the cut--off 
$\gL\ra\infty$ are depicted in figure \ref{localPosibilities}. 
\begin{figure}[!ht]
{\small 
\begin{center}
\renewcommand{\arraystretch}{1.5}
\begin{tabular}{l | c | c }
$q_b \neq 0$ & $\gx_0'' q_b \geq 0$ & $\gx_0'' q_b < 0$ 
\\[2ex] \hline &&\\[-1ex]
\raisebox{10mm}{$\arry{c}{\gx_0 q_b < 0 \\ (\gx_\gp q_b >0)}$}~~ & 
~~~
\raisebox{-2mm}{\scalebox{.7}{\mbox{\input{loc0.pstex_t}}}}
~~~
& 
~~~
\raisebox{-2mm}{\scalebox{.7}{\mbox{\input{near0.pstex_t}}}}
~~~
\\[-1ex] &&\\ \hline &&\\[-1ex]
\raisebox{10mm}{$\arry{c}{\gx_\gp q_b < 0 \\ (\gx_0 q_b > 0)}$} & 
\raisebox{-2mm}{\scalebox{.7}{\mbox{\input{locpi.pstex_t}}}}
& 
\raisebox{-2mm}{\scalebox{.7}{\mbox{\input{nearpi.pstex_t}}}}
 \\[-1ex] &&\\ \hline &&\\[-1ex]
\raisebox{10mm}{$\arry{c}{\gx_0= \gx_\gp = 0 \\ \gx_0'' \neq 0}$} & 
\raisebox{-2mm}{\scalebox{.7}{\mbox{\input{loc0pi.pstex_t}}}}
& 
\raisebox{-2mm}{\scalebox{.7}{\mbox{\input{cons0pi.pstex_t}}}}
\\[-1ex] &&\\ \hline 
\end{tabular}
\end{center}
}
\caption{This table schematically displays the different shapes of 
a (bulk) zero mode with charge $q_b \neq 0$ when the cut--off 
$\gL$ is taken to be very large.}
\label{localPosibilities}
\end{figure}
When a zero mode becomes localized, it is natural to ask what happens to
the other (massive) states in the KK--towers. Taking into account the 
non--trivial background for $\langle \gF \rangle$ induced by the 
FI--terms, the KK--mass spectrum is given by    
\equ{
(m_n^b)^2 = \frac 14 (g_5 q_b \gx_0)^2 + \frac {n^2}{R^2}, 
\quad 
n \in \Natr.  
\labl{Spectrum}
}
Clearly, in the limit of large cut-off $\gL$ all non-zero mode states
become extremely heavy, and should decouple from the theory.

\section{Orbifold anomalies}

The remainder of this talk is devoted to the subject of anomalies 
that can be associated to orbifold models. First we briefly mention
the parity anomaly on the circle and then turn our attention to the 
gauge anomalies on the orbifold. 
The construction of the orbifold field theory relies on the fact that 
\equ{
\gps(-y) = i \gg^5 \gps(y), \quad
A_{\gm}(-y) = A_{\gm}(y), \quad
A_{5}(-y) = -A_{5}(y), 
\labl{S1parity} 
}
is a symmetry of the theory on the circle which can be modded out, so
as to obtain the orbifold theory. However, one has to be careful since 
this symmetry can be anomalous as was observed in ref. 
\cite{Alvarez-Gaume:1985nf}. This anomaly can be canceled by adding
the parity anomaly counter term 
\(
\gG_{PAC}(A) = 
- \gp i  \int_{S^{4} \!\times\! S^1} \gO_{5}(A),
\labl{parity_counter}
\)
with $\gO_5(A)$ the Chern-Simons $5$-form. (See for example 
ref.\cite{Nakahara:1990th} for the definitions of forms 
$\gO_{2n+2}, \gO_{2n+1}$ and $\gO^1_{2n}$.) But then invariance 
under non--contractible gauge transformations may be lost.
If this happens, it may not be possible to define a consistent quantum
field theory on this orbifold. Here we do not go into detail but just 
state a rule of thumb: if the number of bulk fermions and the sum of
their charges is even, no parity anomaly arises 
\cite{GrootNibbelink:2002xx}. 

Next, we consider gauge anomalies on the orbifold $S^1/\Intr_2$. 
Using a variant of the argumentation of Horava and Witten 
\cite{Horava:1996qa}, we infer that the gauge anomaly of this five
dimensional theory is localized at the fixed points 
\equ{
\gd_\gL \gG(A) = 
N \, \gp i  \int_{{\mathcal M}_5} \left( \gd(y) + \gd(y - \gp R) \right) 
\gO^1_{4|F}(A; \gL) \d y,
}
where $\gG(A)$ denotes the effective action with the fermions 
integrated out. Here  the anomaly is normalized to the fundamental 
representation $F$ to fix the normalization $N$ of the anomaly 
uniquely. (Using a perturbative calculation a similar 
result was obtained in ref.\cite{Arkani-Hamed:2001is}. See also 
refs.\cite{Pilo:2002hu,Barbieri:2002ic} for a discussion on the orbifold 
$S^1/\Intr_2\!\times\!\Intr_2$.)
When there are chiral fermions on the boundaries we may obtain 
additional anomaly contributions: the variation of their effective
action $\gG_I(A)$ reads 
\equ{
\gd_\gL \gG_I(A)
=  N_I\, 2\gp i \left.  \int_{S^{4}} \gO^1_{4 |F}(A; \gL) 
\right|_{I R}.
\labl{brane_gauge_anom}
}
In addition we can allow for a five dimensional Chern-Simons action
$\gG_{CS}(A)$: 
\[
\gd_{\gL} \gG_{CS}(A) = 
 N_{CS}\, \gd_{\gL} \gp i \int_{{\mathcal M}_5} \gO_{5 |F}(A) 
\qquad\qquad\qquad \qquad \quad 
\]
\equ{
\qquad \qquad \qquad \qquad 
= N_{CS}\, \gp i \int_{{\mathcal M}_5}\left( -\gd(y) + \gd(y - \gp R) \right) 
\gO^1_{4 |F}(A; \gL) \d y.
\labl{CS_anom}
}
To have a theory which does not have any anomaly we find the requirements 
\equ{
N_{CS} = N_0 - N_\gp,
\qquad 
N + N_0 + N_\gp = 0. 
}
Notice that the consistency requirement takes the form of a sum rule,
and is determined by the fermionic zero mode spectrum of the bulk and branes
only. Furthermore, a Chern-Simons term is required only if
the anomalies at both branes are not equal to each other.

In the final part of this talk we combine the localization effects and 
the discussion on the gauge anomalies. One may define a
``massive'' anomaly as the anomaly of the bulk fields minus the bulk
zero mode. Now if due to FI--terms the zero mode gets localized at a
brane, the massive anomaly equals the variation of the 
Chern--Simons term. Pictorially this may be represented as 
\equ{
\raisebox{-7mm}{
{\scalebox{.45}{\mbox{\input{CS.pstex_t}}}}
}
{\ =\ } 
\raisebox{-7mm}{
{\scalebox{.45}{\mbox{\input{bulk_anom.pstex_t}}}} 
}
{\ -\ } 
\raisebox{-7mm}{
{\scalebox{.45}{\mbox{\input{local0_anom.pstex_t}}}}
}
{.} 
\labl{MassiveAnomaly}
}
This shows that the localized bulk zero mode cancels the anomaly of
the brane chiral fermions; while the heavy stuff from the four
dimensional effective field theory point of view (the anomaly due to 
the massive KK--states and the Chern--Simons variation) cancel among 
themselves, leaving no trace in the zero mode four dimensional theory.

\section*{Acknowledgments}

Work supported in part by the European Community's Human Potential
Programme under contracts HPRN--CT--2000--00131 Quantum Spacetime,
HPRN--CT--2000--00148 Physics Across the Present Energy Frontier
and HPRN--CT--2000--00152 Supersymmetry and the Early Universe.
SGN was partially supported by CITA and NSERC.

\end{document}